\documentclass[superscriptaddress, twocolumn,
 amsmath,amssymb,
 aps,
]{revtex4-1}

\usepackage{graphicx}
\usepackage{dcolumn}
\usepackage{bm}

\newcommand{\abs}[1]{\left\vert#1\right\vert}

\newcommand{\mb}[1]{\mathbf{#1}}
\newcommand{\mrm}[1]{\mathrm{#1}}

\newcommand{\mc}[1]{\mathcal{#1}}

\newcommand{\ev}[1]{\mrm{E}[{#1}]}

\newcommand{\evbig}[1]{\mrm{E}\bigg[{#1}\bigg]}
\newcommand{\psd}[1]{S_{#1}(\omega)}
\newcommand{\psdts}[1]{S^{TS}_{#1#1}(\omega)}

\newcommand{\eacts}[1]{R^{TS}_{#1#1}(\Delta t)}

\newcommand{\mvar}[2]{\sigma^2_{y#2}(#1)}

\newcommand{\svar}[2]{\sigma^2_{y#2}[#1]}

\newcommand{\avar}[2]{{^A}\sigma^{2}_{y#2}(#1)}
\newcommand{\y}[1]{\bar{y}_{#1}}
\newcommand{\ylo}[1]{\bar{y}^{LO}_{#1}}
\newcommand{\yllo}[1]{\bar{y}^{LLO}_{#1}}
\newcommand{\gb}[1]{\bar{g}_{#1}}
\newcommand{\ft}[1]{\mc{F}\{#1\}}
\newcommand{\ift}[1]{\mc{F}^{-1}\{#1\}}
\newcommand{\tf}[1]{\abs{G_{#1}(\omega)}^2}

\newcommand{\atf}[1]{\abs{{^{A}_{#1}}G(\omega)}^{2}}
\newcommand{\pairtf}[2]{G_{#1,#2}^2(\omega)}
\newcommand{\covtf}{G^2(\omega)}

\newcommand{\cov}[2]{\sigma(#1,#2)}
\newcommand{\uramsey}{T_{R}}
\newcommand{\ramsey}[1]{T_{R}^{(#1)}}

\newcommand{\floor}[1]{\lfloor#1\rfloor}

\usepackage{color}
\definecolor{BLACK}{gray}{0}
\definecolor{WHITE}{gray}{1}
\definecolor{RED}{rgb}{1,0,0}
\definecolor{GREEN}{rgb}{0,1,0}
\definecolor{BLUE}{rgb}{0,0,1}
\definecolor{CYAN}{cmyk}{1,0,0,0}
\definecolor{MAGENTA}{cmyk}{0,1,0,0}
\definecolor{YELLOW}{cmyk}{0,0,1,0}
 
\usepackage[
colorlinks=true,linkcolor=red,citecolor=green,urlcolor=blue,
pdfstartview=FitH,
bookmarksopen=true,bookmarksopenlevel=0,
plainpages=false,pdfpagelabels,
pagebackref=false,
pdftoolbar=false]{hyperref}

\begin{document}

\preprint{APS/123-QED}

\title{Analytically exploiting noise correlations inside the feedback loop to improve locked-oscillator performance}

\author{J. Sastrawan}
\affiliation{ARC Centre for Engineered Quantum Systems, School of Physics, The
University of Sydney, NSW 2006 Australia\\ \emph{and} National Measurement Institute, West Lindfield, NSW 2070 Australia}
\author{C. Jones}
\affiliation{ARC Centre for Engineered Quantum Systems, School of Physics, The
University of Sydney, NSW 2006 Australia\\ \emph{and} National Measurement Institute, West Lindfield, NSW 2070 Australia}

\author{I. Akhalwaya}
\author{H. Uys}
\affiliation{National Laser Centre, Council for Scientific and Industrial Research, Pretoria, South Africa}
\author{M.J. Biercuk}%
\email{michael.biercuk@sydney.edu.au}
\affiliation{ARC Centre for Engineered Quantum Systems, School of Physics, The
University of Sydney, NSW 2006 Australia\\ \emph{and} National Measurement Institute, West Lindfield, NSW 2070 Australia}
\date{\today}%

\date{\today}

\begin{abstract}
We introduce concepts from optimal estimation to the stabilization of precision frequency standards limited by noisy local oscillators.  We develop a theoretical framework casting various measures for frequency standard variance in terms of frequency-domain transfer functions, capturing the effects of feedback stabilization via a time-series of Ramsey measurements.  Using this framework we introduce a novel optimized hybrid \emph{predictive feedforward} measurement protocol which employs results from multiple past measurements and transfer-function-based calculations of measurement covariance to improve the accuracy of corrections within the feedback loop.  In the presence of common non-Markovian noise processes these measurements will be correlated in a calculable manner, providing a means to capture the stochastic evolution of the LO frequency during the measurement cycle.   We present analytic calculations and numerical simulations of oscillator performance under competing feedback schemes and demonstrate benefits in both correction accuracy and long-term oscillator stability using hybrid feedforward.  Simulations verify that in the presence of uncompensated dead time and noise with significant spectral weight near the inverse cycle time predictive feedforward outperforms traditional feedback, providing a path towards developing a new class of stabilization ``software'' routines for frequency standards limited by noisy local oscillators.


\end{abstract}

\pacs{Valid PACS appear here}
\maketitle

High-performance passive frequency standards play a major role in technological applications such as network synchronization and GPS~\cite{katori2011} as well as many fields of physical inquiry, including radioastronomy (very-long-baseline interferometry) \cite{nand2011}, tests of general relativity \cite{chou2010o}, and particle physics \cite{blatt2008}.   Atomic clocks exploiting the stability of Cs~\cite{guena2012,bipm2006,sullivan2001,audoin2001} or other atomic references ~\cite{fisk1997, hinkley2013,huntemann2012,chou2010f,rosenband2008} to stabilize an oscillator are known as the most precise timekeeping devices available, but constant performance gains are sought for technical and scientific applications.

In many settings, such as miniaturized deployable frequency standards or in GPS-denied environments, a major performance limitation aries from the quality of the local oscillator (LO) that probes and is locked to the atomic transition.  The LO frequency may evolve randomly in time due to intrinsic noise processes in the underlying hardware~\cite{hinkley2013,huntemann2012}, leading to time-varying deviations of the LO frequency from that of the stable atomic reference.  These instabilities are \emph{partially} compensated through use of a feedback protocol designed to transfer the stability of the reference to the LO, but their effects cannot be mitigated completely.  

Early work characterizing the so-called Dick effect~\cite{dick1987} demonstrated that no matter how good the reference becomes, LO noise will still produce residual instabilities in the locked LO (LLO) through the feedback protocol itself.  The dominant mechanism for this is evolution of the LO's frequency on timescales rapid compared with the shortest measurement and feedback cycle.  Major contributors to this phenomenology relate to the presence of uncompensated LO evolution during initialization and readout stages of the measurement cycle (dead time), as well as aliasing of LO noise at harmonics of the feedback-loop period -- the Dick effect \cite{dick1987,jiang2011,takamoto2011}.  Accordingly, significant research focus in the frequency standards community has been placed on improving LO performance, using e.g. ultra-low-phase-noise cryogenic sapphire oscillators or similar \cite{hartnett2010,hagemann2013}, with concomitant increases in hardware infrastructure requirements and complexity.   Other approaches to mitigating the impact of LO instabilities involve significant modification of the relevant reference hardware, for instance employing multiple atomic references~\cite{hinkley2013,biedermann2013}.
  
In this Manuscript we devise and analyze a method by which both the accuracy of the LLO relative to the atomic reference, and the stability of the composite passive frequency standard, can be improved without the need for hardware modification.  We develop new analytic tools casting time-domain statistical measures of frequency-standard performance in terms of analytically calculable transfer functions~\cite{TransferFunction}, exploiting recent related work in quantum information~\cite{KurizkiPRL2001, GreenPRL2012, GreenNJP2013, SoareNatPhys2014}.  This approach reveals opportunities to exploit non-Markovianity in the dynamics of LO frequency fluctuations in order to improve feedback stabilization by bringing optimal estimation inside the feedback loop of the LO.

Our method expresses the properties of the LLO in terms of the statistics of the unlocked LO at different times as well as correlations between those measurements.  We present the relevant transfer functions for time-series measurements of arbitrary-duration Ramsey measurements, and introduce the pair-covariance transfer function explicitly capturing correlations between measurement outcomes at different times.  Thus, given statistical knowledge of the LO noise characteristics, we craft a new form of hybrid feedforward stabilization incorporating the results of an arbitrary number of past measurements with variable duration to calculate an improved correction to the LO.  This approach shares concepts with techniques of optimal estimation~\cite{Stengel} commonly used in engineering to predict the evolution of a dynamical system -- here the noisy LO.

 In cases where dead time is significant and there is substantial uncompensated LO evolution, we use numerical simulations to show that this approach allows corrections of improved accuracy to be applied to the LO.  Simulations demonstrate that long-term stability of the LLO is improved through a moving-average correction scheme, where corrections are made based on weighting values determined analytically in the same hybrid feedforward approach.  The method described here is a technology-independent \emph{software-oriented} approach to improving the performance of frequency standards derived from locked local oscillators. It may be freely used in conjunction with hardware modifications targeted at reducing the same limitations identified, such as interleaving the cycles of two clocks to reduce dead time \cite{hinkley2013,biedermann2013}.

The remainder of this manuscript is organized as follows.  In Section~\ref{Sec:LONoise} we provide an analytic description of the deleterious effects of LO noise on frequency standards, introducing the relevant metrics for performance of interest.  This includes presentation of novel analytic expressions explicitly capturing the effects of feedback stabilization on the aggregate system performance through a recursive formulation.  Section~\ref{Sec:Fourier} demonstrates how to convert these time-domain statistical measures of frequency-standard performance to the Fourier domain, introducing both transfer functions for individual measurements and the pair-covariance transfer function capturing the correlations between arbitrary-duration Ramsey measurements conducted at arbitrary times.  We then exploit these tools in Sec.~\ref{Sec:HFF} in order to devise a new hybrid-feedforward correction scheme similar in spirit to concepts from optimal estimation in order to maximize the accuracy of corrections applied to the LO.  We demonstrate improvements in correction accuracy and LLO stability via this approach using numerical simulations with realistic LO noise power spectra.  Finally, we conclude with a summary and discussion in Sec.~\ref{Sec:Conclusion}.

\section{\label{Sec:LONoise} The effect of local oscillator noise on frequency standard stability}

\begin{figure} [tp]
\includegraphics[width=0.48\textwidth]{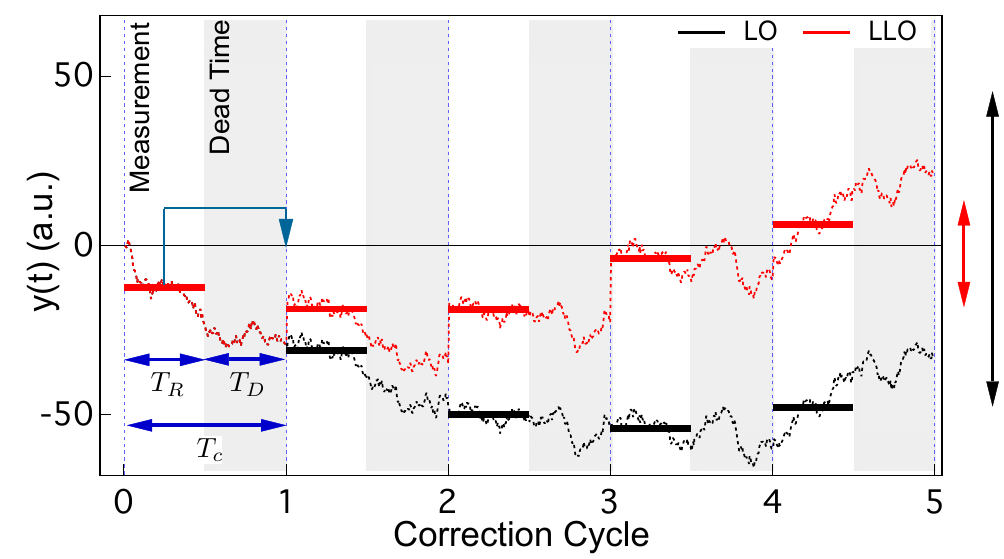}
\caption{\label{fig:traces}Effect of LO noise on the performance of a locked oscillator.  Simulated evolution for a noisy LO, unlocked (black) and locked with traditional feedback (red).  The dotted horizontal bars indicate the measurement outcomes (\emph{samples}) over each cycle, $\y{k}$, which are applied as correction at the end of the cycle, indicated by the bent arrow in the first cycle.   Measurement period of duration $T_{R}$ (white background) is followed by dead time with duration $T_{D}$ (grey background).  Total cycle time $T_{c}=T_D + \uramsey$, and here we represent a 50$\%$ duty factor, $d$.  Undetected evolution of the LO during the dead time leads corrections to incompletely cancel frequency offsets at the time of correction.  The arrows on the far right schematically indicate how locking reduces the variance of $y(t)$ though it does not eliminate it.}
\end{figure}

Our primary objective is to suppress the impact of LO noise on the ultimate performance of the \emph{locked} LO, which is stabilized to an (in general atomic) reference.  Accordingly, throughout this analysis we do not consider systematic shifts or uncertainties in the reference and explicitly assume that the reference is perfect.  

We represent the fractional frequency offset of the LO relative to an ontologically perfect reference $y(t) \equiv (\nu(t) - \nu_0)/\nu_0$, where $\nu_0$ is the reference frequency and $\nu(t)$ is the LO frequency.  This limit provides a reasonable approximation to the performance of many deployable frequency standards where LO stability is far worse than that of the associated atomic reference.

\subsection{Time-domain description of Ramsey measurements and feedback stabilization }
In such a setting, Ramsey spectroscopy provides a means to determine the fractional frequency offset of the LO relative to the reference over a period $\uramsey{}$.  Point-like realisations of the stochastic process $y(t)$ cannot be obtained experimentally; instead, the LO frequency error produces integrated \emph{samples}, denoted $\bar{y}_k$ and indexed in time by $k$:

\begin{align}\label{eq:yk}
\y{k} &\equiv \frac{1}{\ramsey{k}} \int_{t^s_k}^{t^e_k} y(t) g(t - t^s_k) dt
\end{align}

\noindent where $\ramsey{k} \equiv t^e_k - t^s_k$, $[t^s_k,t^e_k]$ is the time interval over which the $k$th sample is taken, and $g(t)$ is a \emph{sensitivity function} capturing the extent to which LO fluctuations at some instant $t$ contribute to the measured outcome for that sample~\cite{rutman1978}. The range of $g(t)$ is $[0,1]$ and its domain is $t\in[0,\ramsey{k}]$. The ideal case is the rectangular window case, where

\begin{align}
g(t) = \begin{cases}
1 & \text{ for } t \in[0,\ramsey{k}] \\
0 & \text{ otherwise }
\end{cases}
\end{align}

\noindent in which case $\y{k}$ reduces to the time-average of $y(t)$ over the interval $[t^s_k,t^e_k]$.  

In traditional feedback stabilization, the samples, $\y{k}$, are used to determine corrections to be applied to the LO in order to reduce frequency differences from the reference (Fig.~\ref{fig:traces}).  Consider the trajectory of the same frequency noise realisation $y(t)$ in the cases of no correction, $y^{LO}(t)$ and correction, $y^{LLO}$(t). The relation between these two cases of $y(t)$ is

\begin{align}\label{eq:diff}
y^{LLO}(t) &= y^{LO}(t) + \sum_{k=1}^{n} C_k
\end{align}

\noindent where $C_k$ refers to the value of the $k$th frequency correction applied to the LO, $n$ of which have occurred before time $t$. 

Under traditional feedback stabilization, each correction is directly proportional to the immediately preceding measurement outcome: $C_k = w_k \yllo{k}$, where $w_k$ is correction gain. Since $\yllo{k}$ is calculated by convolving $y^{LLO}(t)$ with a sensitivity function pertaining to the measurement parameters, (\ref{eq:diff}) is a recursive equation in general. It is possible to cancel all but one of the recursive terms by setting the correction gain equal to the inverse of the average sensitivity $\gb{k} \equiv \int_0^{\ramsey{k}} g(t)/\ramsey{k} dt$ of the preceding measurement, i.e. $w_k = -\gb{k}^{-1}$, where the minus sign indicates negative feedback. With this constraint we can write

\begin{align}\label{eq:nonrec}
\yllo{k} &= \ylo{k} - \frac{\gb{k}}{\gb{k-1}}\ylo{k-1}
\end{align}

\noindent and for a Ramsey interrogation and measurement with negligibly short pulses, $\gb{k} = 1$.  Applying feedback corrections sequentially after measurements is able to effectively reduce $y(t)$ over many cycles, improving long-term stability.

In the limit of a static offset, a single (perfect) correction will set the frequency offset error of the LLO to zero; however, such perfect correction is in general not achieved.  The primary reason for this in the limit of perfect measurements and corrections is \emph{dynamic evolution} of the LO on timescales rapid compared to the measurements which cannot be fully compensated by the feedback loop.  

In Fig.~\ref{fig:traces} we demonstrate how evolution of the LO frequency during $\uramsey{}$ leads the feedback protocol to incompletely correct the offset $y(t)$.   From the formalism presented above we see that incomplete feedback arises because the corrections are based only on the \emph{average} value of the frequency offset as measured over the $k$th period, $\y{k}$ (horizontal solid lines in Fig.~\ref{fig:traces}), rather than the instantaneous value of the LO frequency offset at the time of correction (here the end of a cycle) \emph{which cannot be known}.  The difference between these two values leads to incomplete compensation of time-varying frequency offsets, and hence residual fractional instability in the quantity $y^{(LLO)}(t)$.


The impact of these effects on the ultimate stability of the LLO is exacerbated in circumstances where there is nonzero \emph{dead time}, $T_{D}$, during which the LO may evolve, but this evolution is not captured by a measurement.  Dead time arises due to e.g. the need to reinitialize the reference between measurements, or perform classical processing of the measurement outcome before a correction can be applied.  

The net impact of this uncompensated evolution is a reduction in the long-term stability of the locked local oscillator.  We now move on to describe the relevant quantitative metrics for LLO \emph{variance} in both free-running and feedback-locked settings.


\subsection{Measures of frequency standard stability for unlocked and locked LOs}
The performance of the frequency standard is statistically characterized by various time-domain measures capturing the evolution of LO frequency as a function of time.

The variance of $\y{k}$, denoted $\mvar{k}{}$ and often called \emph{true variance} \cite{rutman1978} is,
\begin{align}
\mvar{k}{} &=\mrm{Var}[\y{k}]=\left( \langle \y{k}^2\rangle - \langle \y{k}\rangle^2\right)\to\ev{\y{k}^2} \\
&= \evbig{\bigg(\frac{1}{\ramsey{k}} \int_{t^s_k}^{t^e_k} y(t) g(t - t^s_k) dt \bigg)^2} 
\end{align}

\noindent where in the first line we assume that the true variance is simply equal to the expected value of $\y{k}^2$, since $y(t)$ is assumed to be a zero-mean process.  The true variance captures the spread of measurement outcomes due to different noise realizations in a single timestep.  However, in a measurement context one does not have immediate access to an infinite ensemble of noise realizations, but rather a single series of measurement outcomes conducted sequentially over a single noise realization.  As a result we rely on a measure more conducive to this setting, the \emph{sample variance} 

\begin{align}
\svar{N}{} = \frac{1}{N-1} \sum_{k=1}^{N} (\y{k} - \frac{1}{N}\sum_{l=1}^{N} \y{l})^2
\end{align}

\noindent for $N$ sequential finite-duration measurements $\{\y{k}\}$~\cite{rutman1978}. 

In this work we will rely on such measures of frequency stability, rather than the more commonly employed Allan variance, in line with recent experiments~\cite{Bergquist_Hg}.  The Allan variance is calculated by finding the variance of the difference between consecutive pairs of measurement outcomes:

\begin{align}\label{eq:allan}
\avar{y}{}&= \frac{1}{2}\langle (\y{k+1} - \y{k})^2 \rangle
\end{align}

\noindent where $\y{k}$ is the $k$th measurement outcome and $\langle\cdots\rangle$ may indicate a time average or an ensemble average, depending on whether $y(t)$ is assumed to be ergodic.  Our decision to avoid the Allan variance is deliberate, as its form -- effectively a moving average -- specifically \emph{masks} the effect of LO noise components with long correlation times.  In fact the Allan variance is employed by the community in part because it does not diverge at long integration times $\tau$ due to LO drifts, as would the sample or true variance \cite{nist1990,barnes1971,rutman1978, Greenhall1999}.  In the limit where the stability of a frequency reference is dominated by LO noise (and the reference can be treated as perfect) this approach gives physically meaningful results.

The standard measures for oscillator performance consider either a free-running LO or provide a means only to statistically characterize measurement outcomes under black-box conditions.  We may derive explicit analytic forms for different measurements of variance in the presence of feedback locking in order to provide insights into opportunities to improve net LLO performance through modification of the stabilization protocol.

 


We write time-domain expressions for variance using the relevant definitions provided above and the link between corrections in feedback and the history of the LLO's evolution.  For the true variance we substitute Eq.~\ref{eq:nonrec} to find
\begin{widetext}
\begin{align}
\mvar{k}{LLO} &= \mrm{Var}[\yllo{k}] \\
&= \mvar{k}{LO} + \bigg(\frac{\gb{k}}{\gb{k-1}}\bigg)^2 \mvar{k-1}{LO} - \frac{2\gb{k}}{\gb{k-1}}\cov{\ylo{k-1}}{\ylo{k}}
\end{align}

\noindent and calculate the expected value of the LLO sample variance in a similar manner using Eq.~\ref{eq:diff}

\begin{align}
\ev{\svar{N}{LLO}} &= \frac{1}{N-1} \sum_{k=1}^{N}\bigg\{ \bigg(\mvar{k}{LO} + \gb{k}^2 \sum_{r=1}^{k-1}\sum_{s=1}^{k-1} \cov{C_r}{C_s} - 2\gb{k}\sum_{u=1}^{k-1} \cov{\ylo{k}}{C_u}\bigg) \nonumber\\
&+ \frac{1}{N^2} \sum_{p=1}^{N}\sum_{q=1}^{N} \bigg( \cov{\ylo{p}}{\ylo{q}} + \gb{p}\gb{q}\sum_{w=1}^{p-1}\sum_{x=1}^{q-1} \cov{C_x}{C_y} \bigg)
- \frac{2}{N} \sum_{l=1}^{N} \bigg(\cov{\ylo{k}}{\ylo{l}} + \gb{k}\gb{l}\sum_{y=1}^{k-1}\sum_{z=1}^{l-1} \cov{C_y}{C_z} \bigg)\bigg\}
\end{align}

\end{widetext}

We see that the characteristics of the locked LO can be expressed in terms of the unlocked LO and the covariance  \emph{covariance} between two quantities, $\cov{x}{y}$, capturing correlations between them.  This may include the covariance of different measurement outcomes on the LO, or different corrections applied to the LO.  It is this observation -- that we may express relevant statistical quantities surrounding the performance of locked local oscillators in terms of measurement covariances -- that will provide a path towards the development of new stabilization routines exploiting temporal correlations in the LO noise (and hence measurement outcomes).

\section{\label{Sec:Fourier}Performance measures for frequency standards in the fourier domain}

We require an efficient theoretical framework in which to capture these effects, and hence transition to the frequency domain, making use of the power spectral density of the LO, $\psd{y}$, in order to characterize average performance over a hypothetical statistical ensemble. In this description residual LLO instability persists because the feedback is insensitive to LO noise at high frequencies relative to the inverse measurement time.  Additional instability due to the Dick effect comes from aliasing of noise at harmonics of the loop bandwidth.

We may analytically calculate the effects of measurement, dead time, and the feedback protocol itself on frequency standard performance in the frequency domain as follows.   Defining a normalised, time-reversed sensitivity function $\bar{g}(t^m_k - t) = g(t - t^s_k)/\ramsey{k}$, where $g(t)$ is assumed to be time-reversal symmetric about $t^m_k$, the midpoint of $[t^s_k,t^e_k]$, we can express, for instance, the true variance as a convolution $\mvar{k}{} = \evbig{\bigg( \int_{-\infty}^{\infty} y(t) \bar{g}(t^m_k - t) dt \bigg)^2}$.  Expanding this expression gives

\begin{align}
\mvar{k}{}&= \int_{-\infty}^{\infty}\int_{-\infty}^{\infty} \ev{y(t) y(t')} \bar{g}(t^m_k - t)\bar{g}(t^m_k - t') dt' dt \\
&= \int_{-\infty}^{\infty} \int_{-\infty}^{\infty} \eacts{y} \bar{g}(t^m_k - t)\bar{g}(t^m_k - t') dt' dt 
\end{align}

\noindent where $\eacts{y}$ is the two-sided autocorrelation function and $\Delta t \equiv t' - t$.  Using the Wiener-Khinchin theorem we write $\eacts{y} = \ift{\psdts{y}}$, relating the autocorrelation function to the Fourier transform of the power spectral density of the LO noise.  Defining the Fourier transform of $\bar{g}(t^m_k - t)$:

\begin{align}\label{eq:tf}
G_k(\omega) &\equiv \int_{-\infty}^{\infty}\bar{g}(t^m_k - t) e^{i\omega t} dt
\end{align}
\noindent We may then express the true variance


\begin{align}
\mvar{k}{}&= \frac{1}{2\pi} \int_{0}^{\infty} \psd{y} \abs{G_k(\omega)}^2d\omega \label{eq:overlap}
\end{align}

\noindent where the substitution of the one-sided PSD $\psd{y}$ is possible because $\abs{G_k(\omega)}^2$ is even. This result is similar to the convolution theorem, which states that $\ft{f \star g} = \ft{f} \cdot \ft{g}$, where $\star$ denotes a convolution and $f$ and $g$ are Fourier-invertible functions.

Here $\tf{k}$ is called the \emph{transfer function} for the $k$th sample, describing the spectral properties of the measurement protocol itself. For measurements performed using Ramsey interrogation with $\pi/2$ pulses of negligible duration and zero dead time, the transfer function has a sinc-squared analytic form $\tf{k} = (\sin{(\omega \ramsey{k}/2)}/(\omega \ramsey{k} /2))^2$.  This framework has recently seen broad adoption in the quantum information community where time-varying dephasing noise is a major concern for the stability of quantum bits~\cite{KurizkiPRL2001, UhrigPRL2007, CywinskiPRB2008, BiercukNature2009, BiercukJPB2011, GreenPRL2012, GreenNJP2013, SoareNatPhys2014}. 

Recalling that statistical measures of LLO variance rely not only on expressions for the true variance over noise ensembles, but also of covariances between measurements or corrections, we must equivalently express the covariance in terms of transfer functions.  Using the identity $\sigma^2(A\pm B) = \sigma^2(A) + \sigma^2(B) \pm 2\sigma(A,B)$, we define a sum and a difference sensitivity function: $g^+_{k,l}(t)$ and $g^-_{k,l}(t)$, with respect to two measurements indexed $k$ and $l$. These expressions are general functions of time with two regions of high sensitivity corresponding to the individual measurement periods.

\begin{align}
g^{\pm}_{k,l}(t) &\equiv \begin{cases} g(t-t^s_k) \text{, for } t \in [t_{k}^s,t_k^e] \\
\pm g(t-t^s_l)  \text{, for } t \in [t_l^s,t_l^e] \\
0 \text{, otherwise}
\end{cases}
\end{align}

\noindent These time-domain sum and difference sensitivity functions have their corresponding frequency-domain transfer functions, defined as their Fourier transforms normalised by $\ramsey{k,l}$:

\begin{align}\label{Eq:AppPairTF}
G_{k,l}^{\pm}(\omega) \equiv \int_{-\infty}^{\infty} \bigg(\frac{g(t^m_k - t)}{\ramsey{k}} \pm \frac{g(t^m_l - t)}{\ramsey{l}}\bigg) e^{i\omega t} dt
\end{align}

\noindent Substituting this and the form of the true variance (\ref{eq:overlap}) into the variance identity above and rearranging terms gives the covariance of the two measurement outcomes

\begin{align}
\cov{\y{k}}{\y{l}} &=  \frac{1}{2\pi}\int_{0}^{\infty} \frac{\psd{y}}{4}\bigg(\abs{G_{k,l}^{+}(\omega)}^2 - \abs{G_{k,l}^{-}(\omega)}^2\bigg)  d\omega\\
&\equiv \frac{1}{2\pi}\int_{0}^{\infty} \psd{y} \pairtf{k}{l} d\omega
\end{align}

\noindent whereby $\pairtf{k}{l}$ is defined to be the \emph{pair covariance transfer function}.  For the case of flat-top Ramsey measurements over the intervals $[t^{s}_{k,l},t^{e}_{k,l}]$ this term takes the form
\begin{align}\label{eq:pairtf}
\pairtf{k}{l} &= (\omega^2 \ramsey{k} \ramsey{l})^{-1}\Big[\cos{(\omega(t^s_l - t^s_k))} + \cos{(\omega(t^e_l - t^e_k))} \nonumber\\
&~~~ - \cos{(\omega(t^e_l - t^s_k))} - \cos{(\omega(t^s_l - t^e_k))} \Big].
\end{align}
\noindent This is a generalization of the transfer function previously derived for the special case of periodic, equal-duration Ramsey interrogations~\cite{rutman1978,barnes1971}, and allows effective estimation of $y(t)$ for any $t$ and for any set of measured samples $\mb{\y{\mathnormal{k}}}$. 

We thus see that this approach allows expression of time-domain LO variances as overlap integrals between $\psd{y}$ and the transfer functions capturing the effects of the measurement and feedback protocol, including correlations between measurements or corrections in time.  Through this formalism we may incorporate arbitrary measurement protocols (e.g. arbitrary and dynamic Ramsey periods and dead times): the underlying physics of e.g. changing linewidth of the measurement is explicitly captured through the form and implicit time-dependence of the transfer function used to characterize the measurement protocol.


\section{\label{Sec:HFF}Exploiting noise correlations to improve feedback stabilization}
Recasting variance metrics for the stability of LOs in terms of transfer functions is particularly powerful because it provides a path to craft new measurement feedback protocols designed to reduce residual variance measures for the LLO by modifying the protocol's spectral response.  Our key insight is that the non-Markovianity of dominant noise processes in typical LOs -- captured through the low-frequency bias in $\psd{y}$~\cite{barnes1971,rutman1978} -- implies the presence of temporal correlations in $y(t)$ that may be exploited to improve feedback stabilization.  These correlations are captured in the set of measurement outcomes  $\mb{\y{k}}$; accordingly \emph{future evolution} of $y(t)$ may be predicted based on a past set of measurements within $\mb{\y{k}}$, so long as the past measurements and point of prediction fall within the characteristic \emph{correlation time} for the LO noise given by $\psd{y}$.  This approach provides a direct means to account for LO evolution that is normally not compensated during \emph{dead time} in the measurement process.

\subsection{Optimal estimator for corrections}
The formal basis of our analytic approach, in summary, is to calculate a covariance matrix in the frequency domain via transfer functions to capture the relative correlations between sequential measurement outcomes of an LLO, and use this matrix to derive a linear \emph{predictor} of the LLO frequency offset at the moment of correction. Under appropriate conditions this predictor provides a correction with higher accuracy than that derived from a single measurement, allowing us to improve the ultimate performance of the LLO.  Since the predictor is found using information from previous measurements (feedback) and a priori statistical knowledge of the LO noise to \emph{predict} the evolution of the LO (feedforward), we call the scheme \emph{hybrid feedforward}.  

This approach shares common objectives with application of optimal control techniques such as Kalman filtering in the production of composite frequency standards from an ensemble of physical clocks~\cite{Greenhall2003}, or in compensating for deterministic frequency shifts due to e.g. aging or changes in the ambient temperature of a clock~\cite{Penrod, Kalman_Clock}.  The primary advance of this work is the insight that \emph{stochastic} evolution of the LO can be predicted and compensated using optimal control protocols \emph{inside the feedback loop}.



In hybrid feedforward, results from a set of $n$ past measurements are linearly combined with weighting coefficients $\mb{c}_k$ optimized such that the $k$th correction, $C_k$, provides maximum correlation to $y(t^c_k)$ at the instant of correction $t^c_k$ (Fig.~\ref{fig:traces}c).   Assuming that the LO noise is Gaussian, the optimal least minimum mean squares estimator (MMSE) is linear, and the optimal value of the correction is given by $C_k = \mb{c}_k \cdot\mb{\y{\mathnormal{k}}}$: the dot product of a set of correlation coefficients $\mb{c}_k$ derived from knowledge of $\psd{y}$ and a set of $n$ past measured samples, $\mb{\y{\mathnormal{k}}} = \{\y{k,1}, \cdots,\y{k,n} \}$.  We define an $(n+1)\times(n+1)$ covariance matrix where the $(n+1)$th term represents an ideal zero-duration sample at $t^c_k$ and in the second line we write the covariance matrix in block form:
\begin{align}
\Sigma_k &\equiv \begin{bmatrix}  \cov{\y{k,1}}{\y{k,1}} & \cdots & \cov{\y{k,1}}{y(t^c_k)} \\[1em] 
\cov{\y{k,2}}{\y{k,1}} & \cdots & \cov{\y{k,2}}{y(t^c_k)} \\[1em]
\cdots & \cdots & \cdots  \\[1em]
\cov{y(t^c_k)}{\y{k,1}} & \cdots & \cov{y(t^c_k)}{y(t^c_k)}
\end{bmatrix}\\
&\equiv \begin{bmatrix} \mb{M}_k & \mb{F}_k \\
\mb{F_\mathnormal{k}^T} & \cov{y(t^c_k)}{y(t^c_k)} \\
\end{bmatrix}.
\end{align}
\noindent  In this form the matrix $\mb{M}_k$ describes correlations between measurement outcomes while the vector $\mb{F}_k$ describes correlations between each measurement and the LLO at the time of correction.  The MMSE optimality condition is then fulfilled for 
\begin{align}
\mb{c}_k &= \frac{\mb{F}_k }{\sqrt{\mb{F_\mathnormal{k}^T}\mb{M}_k\mb{F}_k}} \frac{w_k}{2\pi} \int_0^{\infty} \psd{y} d\omega
\end{align}
\noindent where $w_k$ is an overall correction gain.  The covariance matrix elements are calculated as defined above in terms of the LO noise power spectrum.  

In the practical setting of a frequency standard experiment, we wish to improve both the accuracy of each correction, by maximising the correlation between $C_k$ and $y(t^c_k)$, and the long-term stability of the LLO output, captured by the metrics of frequency variance, sample variance, and Allan variance.  

\begin{figure} [bp]
\includegraphics[width=0.48\textwidth]{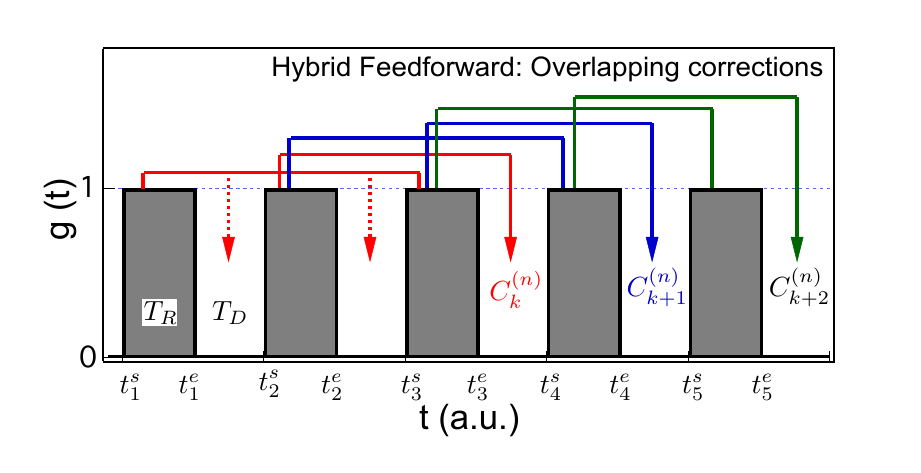}
\caption{\label{Fig:HFF} Schematic diagram of hybrid feedforward with an example protocol using $n=3$.  Start and end times of measurements are defined arbitrarily permitting non-uniform-duration measurements, although measurements are illustrated as uniform for clarity.  Corrections $C^{(n=3)}_k$ are applied in either non-overlapping blocks of three measurements or as a moving average (depicted here).  In the latter case, the covariance matrix must be recalculated to correctly account for any variations in measurement duration.  Dashed red arrows indicate the first corrections performed without full calculation of the covariance matrix.  This effect vanishes for $k>n$.}
\end{figure}

Although the LLO frequency variance under hybrid feedforward for more than a single cycle cannot be expressed in a closed non-recursive form, a consideration of a single cycle can provide a value for $\langle y^{LLO}(t^c_k)^2 \rangle$ in terms of covariance matrix elements.  This in turn provides a metric for the \emph{correction accuracy} for hybrid feedforward, defined as the extent to which a correction brings $y^{LLO}(t)\to0$ at the instant of correction, $t=t^c_k$

\begin{align}\label{Eq:Accuracy}
A_k &\equiv  \frac{\langle y^{LO}(t^c_k)^2 \rangle}{\langle y^{LLO}(t^c_k)^2 \rangle} \\
&= \bigg(1 + w_k^2 - w_k \frac{\abs{\mb{F_k}}^2}{\sqrt{\mb{F^T_k M_k F_k}}} \bigg)^{-1}
\end{align}

We can gain insights into the performance of the correction protocol by considering limiting cases.  For instance, in the limit of white noise with negligible correlations, $\mb{M_k}\to\mathbb{I}$,the identity matrix. In this limit the rightmost term in Eq.~\ref{Eq:Accuracy} reduces to $w_{k}\abs{\mb{F_k}}$, which is small (there are negligible correlations between measurement outcomes and $y(t^{c}_{k})$). In this limit, accuracy $A_k \to 1/(1+ w_k)$, and is maximized by setting $w_{k}=0$ (not performing feedback at all) as corrections are uncorrelated with $y(t^{c}_{k})$.  By contrast with perfect correlations all elements of the covariance matrix take value unity.  Standard feedback works perfectly by selecting unity gain and selecting the number of measurements to be combined, $n=1$, to correct based on a single measurement.  

In intermediate regimes the ensemble-averaged accuracy of the hybrid feedforward correction is determined in by a balance of covariance between elements of $\mb{\y{k}}$ and covariance between $\mb{\y{k}}$ and $y(t^{c}_{k})$, the LO noise at the time of correction.  Achieving correction which improves LLO variance requires setting the term in parentheses in Eq.~\ref{Eq:Accuracy} to less than unity.  This in turn places a condition on the correlations in the system

\begin{align}
\sqrt{\mb{F^T_k M_k F_k}}<\frac{\abs{\mb{F_k}}^2}{w_{k}}
\end{align}

\noindent We can interpret the effect of $\mb{M_k}$ as an effective rotation matrix, reducing the magnitude of the left-hand side of the expression above by effectively maximizing the ``angle'' between $\mb{M_{k}F_{k}}$ and $\mb{F_{k}}$.  While it is unphysical to reduce this to zero based on the limiting cases discussed above, it is possible to appropriately select $k$, based on characteristics of $\psd{y}$ in order to improve correction accuracy.

In all slaved frequency standards we rely on repeated measurements and corrections to provide long-term \emph{stability}, a measure of how the output frequency of the LLO deviates from its mean value over time.  We study this by calculating the sample variance of a time-sequence of measurement outcomes averaged over an ensemble of noise realizations, $\langle\svar{N}{}\rangle$. A ``moving average'' style of hybrid feedforward provides improved long-term stability, as the correction $C_k$ will depend on the set of measurement outcomes $\mb{\y{\mathnormal{k}}} = \{\y{k-n+1}, \cdots, \y{k} \}$, among which previous corrections have been interleaved, as illustrated in Fig.~\ref{Fig:HFF}.  In this case the covariance matrix must be updated to reflect the action of each correction.  See Appendix for a detailed form of the Sample Variance in the case of this form of stabilization.

\subsection{Numerical Simulations}
In order to test the general performance of hybrid feedforward in different regimes we perform numerical simulations of noisy LOs with user-defined statistical properties, characterized by $\psd{y}$.  We produce a fixed number of LO realizations in the time domain and then use these to calculate measures such as the sample variance over a sequence of ``measurement'' outcomes with user-defined Ramsey measurement times, dead times, and the like. In these calculations we may assume that the LO is free running, experiencing standard feedback, or employing hybrid feedforward, and then take an ensemble average over LO noise realizations.  Our calculations include various noise power spectra, with tunable high-frequency cutoffs, including common `flicker frequency' ($\psd{y} \propto 1/\omega$), and `random walk frequency' ($\psd{y} \propto 1/\omega^{2}$) noise, as appropriate for experiments incorporating realistic LOs,


Tunability in the hybrid feedforward protocol comes from the selection of $n$, in determining $\{C_k\}$ as well as the selected Ramsey periods, permitting an operator to sample different parts of $\psd{y}$.  As an example, we fix our predictor to consider $n=2$ sequential measurements and permit the Ramsey durations to be varied as optimization parameters. A Nelder-Mead simplex optimization over the measurement durations finds that a hybrid feedforward protocol consisting of a long measurement period followed by a short period maximizes correction accuracy (Fig.~\ref{Fig:Accuracy}). This structure ensures that low-frequency components of $\psd{y}$ are sampled but the measurement sampling the highest frequency noise contributions are maximally correlated with $y(t^{c}_{k})$.   With $\psd{y}\propto 1/\omega$ and $\psd{y}\propto 1/\omega^{2}$ we observe increased accuracy under hybrid feedforward while the rapid fluctuations in $y(t)$ arising from a white power spectrum mitigate the benefits of hybrid feedforward, as expected.  In the parameter ranges we have studied numerically we find that correction accuracy is maximized for $n=2$ to $3$, with diminishing performance for larger $n$.  Again, this is determined by the relevant correlation time of the LO noise.

\begin{figure}
\includegraphics[width=0.48\textwidth]{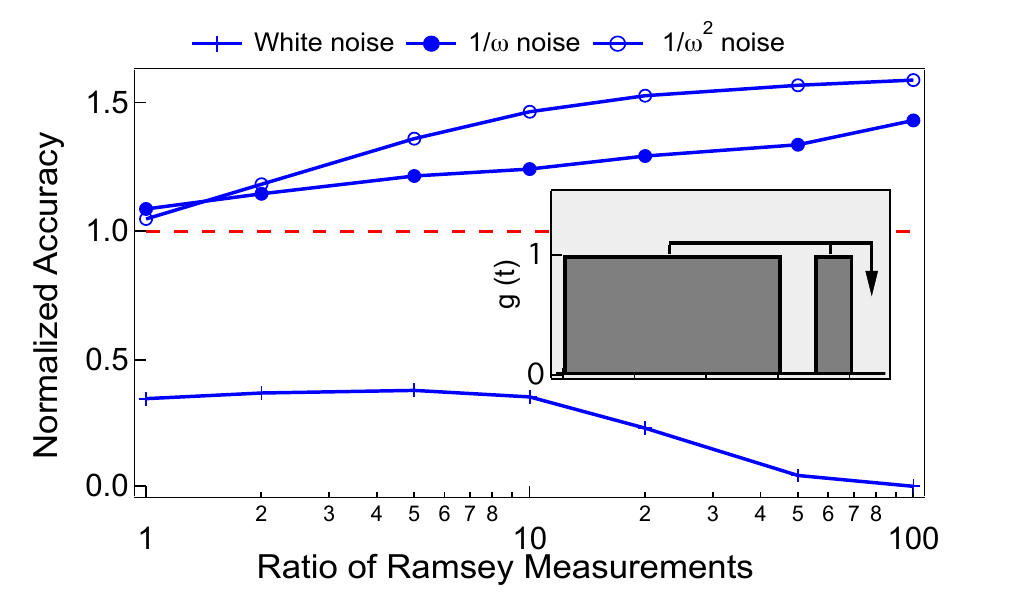}
\caption{\label{Fig:Accuracy}Calculated correction accuracy of the first correction for hybrid feedforward normalized to feedback (accuracy = 1), under different forms of $\psd{y}$ as a function of the ratio of Ramsey periods between the two measurements employed in constructing $C^{(2)}_{k}$.  Correction accuracy for feedback is calculated assuming the minimum Ramsey time; thus for the ratio of Ramsey measurements taking value unity on the $x$-axis, the hybrid feedforward scheme takes twice as long as feedback.  Inset: depiction of the form of $C^{(2)}_{k}$ used in hybrid feedforward, depicting the ``slower'' measurement being performed first.}
\end{figure}


In Fig.~\ref{fig:svar}b we demonstrate the resulting \emph{normalized improvement} in $\langle\svar{N}{}\rangle$ up to $N=100$ measurements, calculated using feedback and hybrid feedforward with $n=2$, and assuming uniform $T_{R}$.  We observe clear improvement (reduction) in $\langle\svar{N}{}\rangle$ through the hybrid feedforward approach, with benefits of order $5-25\%$ of $\langle\svar{N}{}\rangle$ relative performance improvement over standard measurement feedback. We present data for different functional forms of $\psd{y}$, including low-frequency dominated flicker noise ($\propto 1/\omega$), and power spectra ($\propto 1/\omega^{1/2}$) with more significant noise near $T^{-1}_{c}$.  The benefits of our approach are most significant in the long term when high-frequency noise reduces the efficacy of standard feedback.  Notably, because of well known relationships between LO \emph{phase noise} and LO \emph{frequency noise}~\cite{nist1990}, significant high-frequency weight in $\psd{y}$ is commonly encountered.

\begin{figure}
\includegraphics[width=0.48\textwidth]{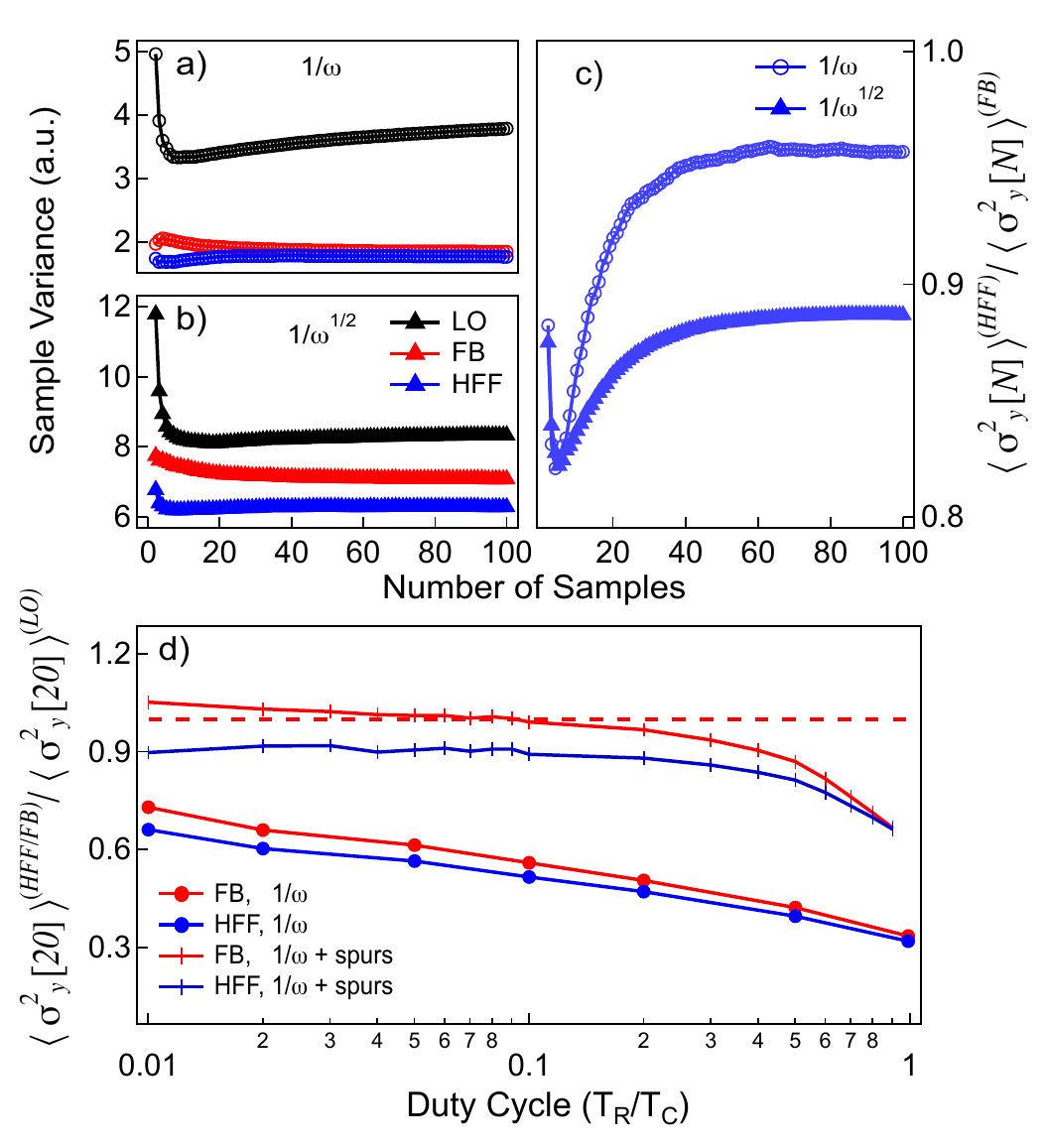}
\caption{\label{fig:svar}(a, b) Calculated sample variance for an unlocked LO, feedback, and hybrid feedforward, as a function of measurement number $N$, for different power spectra (indicated on graphs).  Calculations assume $\psd{y}\propto 1/\omega$,  with a high-frequency cutoff $\omega_{c}/2\pi=100/T_{c}$  and $\psd{y}\propto 1/\omega^{1/2}$ with a cutoff frequency $\omega_{c}/2\pi=1/T_{c}$, demonstrating the importance of high-f noise near $\omega/2\pi=T^{-1}_{c}$. PSDs with different $\omega$-dependences are normalised to have the same value at $\omega_{low} = 1/100T_{c}$.  (c) Normalized sample variance data from panels (a) and (b) presented as the ratio of $\langle\svar{N}{}\rangle^{(HFF)}/\langle\svar{N}{}\rangle^{(FB)}$ in order to demonstrate improvement due to hybrid feedforward (numbers less than unity indicate smaller sample variance under hybrid feedforward).   (d)  Calculated $\langle\svar{N}{}\rangle$ for $N=20$ as a function of duty factor, normalized to the sample variance for the free-running LO.  Data above red dashed line indicate that the standard feedback approach produces instability \emph{larger} than that for the free-running oscillator.  Both data sets assume $\psd{y}\propto 1/\omega$,  with $\omega_{c}/2\pi=100/T_{c}$.  Crosses represent data with ten noise spurs superimposed on $\psd{y}$, starting at $\omega/2\pi=1.15T^{-1}_{c}$, and increasing linearly with step size $0.15T^{-1}_{c}$. }
\end{figure}

 In Fig.~\ref{fig:svar}c we calculate the expectation value of the sample variance at a fixed value of $N=20$ for a LLO stabilized using either traditional feedback or hybrid feedforward.  The sample variances are normalized by that for the free-running LO, meaning that values of this metric less than unity demonstrate improvement due to stabilization, and smaller values indicate better stabilization.  On the horizontal axis we vary the duty factor $d$, defined as the ratio of the interrogation time to total cycle time: $d\equiv  \uramsey/T_c$ from $1\%$ to unity (no dead time), and we compare $\psd{y}\propto 1/\omega$ and $\psd{y}\propto 1/\omega^{1/2}$.  These power spectra are conservative but inspired by typical LO phase noise specifications weighted to enhanced high-frequency content due to the conversion between phase and frequency instability~\cite{nist1990}.
 
 This improvement provided by hybrid feedforward is most marked for low duty factor $d$.  As $d\to1$ the performance of traditional feedback and hybrid feedforward converge, as standard feedback corrections become most effective when dead time is shortest.  However as the dead time increases, and in the presence of $\psd{y}$ with frequency weight near $T_c$, feedback efficacy diminishes due to uncompensated evolution of the LO during the dead time.  

In this regime knowledge of correlations in the noise allows hybrid feedforward to provide metrologically significant gains in stability relative to traditional feedback.   In Fig.~\ref{fig:svar}d we further demonstrate that in the presence of a typical $1/\omega$ power spectrum, the inclusion of noise spurs near $\omega/2\pi=T^{-1}_{c}$ results in certain regimes where standard feedback makes long-term stability \emph{worse} than applying no feedback at all, while feedforward provides useful stabilization. This significant difference arises because even though the noise processes are random, knowledge of the statistical properties of the noise provides a means to effectively model the average dynamical evolution of the system, and accurately predict how the system will evolve in the future.  Exact performance depends sensitively on the form and magnitude of $\psd{y}$, but results demonstrate that systems with high-frequency noise content around $\omega/2\pi \approx T^{-1}_{c}$ benefit significantly from hybrid feedforward.  



\section{\label{Sec:Conclusion}Conclusion}
In summary, we have presented a set of analytical tools describing LLO performance in the frequency domain for arbitrary measurement times, durations, and duty cycles.  We have employed these generalized transfer functions to develop a new software approach to LO feedback stabilization in slaved passive frequency standards, bringing optimal estimation techniques inside the feedback loop.  This technique leverages a series of past measurements and statistical knowledge of the noise to improve the accuracy of feedback corrections and ultimately improve the stability of the slaved LO.  We have validated these theoretical insights using numerical simulations of noisy local oscillators and calculations of relevant stability metrics.  

The results we have presented have not by any means exhausted the space of modifications to clock protocols available using this framework.   For instance we have numerically demonstrated improved correction accuracy using nonuniform-duration $T_{R}$ over a cycle, as well as long-term stability improvement using only the simplest case of uniform $T_{R}$.  These approaches may be combined to produce LLOs with improved accuracy relative to the reference at the time of correction and improved long-term stability.  In cases where the penalty associated with increasing $T_{R}$ is modest (lower high-frequency cutoff), such composite schemes can provide substantial benefits as well, improving both accuracy of correction to the LLO and overall frequency standard stability.  Other expansions may leverage the basic analytic formalism we have introduced; we have introduced the transfer functions, $\tf{}$ and $\pairtf{k}{l}$, but have assumed only the simplest form for the time-domain sensitivity function and fixed overall gain.  However, it is possible to craft a measurement protocol to yield $\tf{}$ that suppresses the dominant spectral features of the LO noise.  We have observed that through such an approach one may reduce the impact of aliasing on clock stabilization, indicating a path for future work on reducing of the so-called Dick limit in precision frequency references.  

In the parameter regimes we have studied the relative performance benefits of the hybrid feedforward approach are of metrological significance - especially considering they may be gained using only ``software'' modification without the need for wholesale changes to the clock hardware.  We believe the approach may find special significance in tight-SWAP (size, weight, and power) applications such as space-based clocks where significantly augmenting LO quality is generally impossible due to system-level limitations.  Overall, we believe that this work indicates clear potential to improve passive frequency standards by incorporation of optimal estimation techniques in the feedback loop itself.  \emph{Note:} While preparing this manuscript we became aware of related work seeking to employ covariance techniques to improve measurements of quantum clocks~\cite{mullan2014}.

\emph{Acknowledgements}: The authors thank H. Ball, D. Hayes, and J. Bergquist for useful discussions.  This work partially supported by Australian Research Council Discovery Project DP130103823, US Army Research Office under Contract Number  W911NF-11-1-0068, and the Lockheed Martin Corporation.

\section*{Appendix}
\subsection*{\label{llovariances}Variances for locked local oscillators with hybrid feedforward}
The standard measures for oscillator performance consider either a free-running LO or provide a means only to statistically characterize measurement outcomes under black-box conditions.  Here we present explicit analytic forms for different measurements of variance in the presence of feedback locking.

The expected value of the LLO sample variance can be found by substituting (\ref{eq:diff}) into the definition of the sample variance, producing a generic expression for traditional feedback (one measurement per correction cycle) and hybrid feedforward (multiple measurements per cycle):

\begin{widetext}
\begin{align}
\ev{\svar{N}{LLO}} &= \frac{1}{N-1} \sum_{k'=1}^{N} \bigg\{\mvar{k'}{LLO} + \frac{1}{N^2} \sum_{p'=1}^N\sum_{q'=1}^N \cov{\yllo{p'}}{\yllo{q'}} - \frac{2}{N} \sum_{l'=1}^N \cov{\yllo{k'}}{\yllo{l'}} \bigg\} \\
&= \frac{1}{N-1} \sum_{k'=1}^{N}\bigg\{ \bigg(\mvar{k'}{LO} + \gb{k'}^2 \sum_{r=1}^{\floor{k'/n}}\sum_{s=1}^{\floor{k'/n}} \cov{C_r}{C_s} - 2\gb{k'}\sum_{u=1}^{\floor{k'/n}} \cov{\ylo{k'}}{C_u}\bigg) \nonumber\\
&+ \frac{1}{N^2} \sum_{p'=1}^{N}\sum_{q'=1}^{N} \cov{\ylo{p'} + \gb{p'}\sum_{p=1}^{\floor{p'/n}}C_p}{\ylo{n} + \gb{q'}\sum_{q=1}^{\floor{q'/n}}C_q} - \frac{2}{N} \sum_{l'=1}^{N} \cov{\ylo{k'} + \gb{k'}\sum_{u=1}^{\floor{k'/n}}C_u}{\ylo{l'} + \gb{l'}\sum_{v=1}^{\floor{l'/n}}C_v}\bigg\} \\
&= \frac{1}{N-1} \sum_{k'=1}^{N}\bigg\{ \bigg(\mvar{k'}{LO} + \gb{k'}^2 \sum_{r=1}^{\floor{k'/n}}\sum_{s=1}^{\floor{k'/n}} \cov{C_r}{C_s} - 2\gb{k'}\sum_{u=1}^{\floor{k'/n}} \cov{\ylo{k'}}{C_u}\bigg) \nonumber\\
&+ \frac{1}{N^2} \sum_{p'=1}^{N}\sum_{q'=1}^{N} \bigg( \cov{\ylo{p'}}{\ylo{q'}} + \gb{p'}\gb{q'}\sum_{p=1}^{\floor{p'/n}}\sum_{q=1}^{\floor{q'/n}} \cov{C_p}{C_q} \bigg) \nonumber\\
&- \frac{2}{N} \sum_{l'=1}^{N} \bigg(\cov{\ylo{k'}}{\ylo{l'}} + \gb{k'}\gb{l'}\sum_{k=1}^{\floor{k'/n}}\sum_{l=1}^{\floor{l'/n}} \cov{C_k}{C_l} \bigg)\bigg\}
\end{align}
\end{widetext}

\noindent where in the case of hybrid feedback, $N$ is defined to be total number of measurements and $n$ is the number of measurements per cycle. The summation signs with unprimed indices are sums over whole cycles (of which there are $\floor{N/n}$) and the primed indices are sums over all $N$ measurements. In general, $\ev{\svar{N}{LLO}}$ contains recursive terms that cannot be concisely expressed in terms of the LO PSD $\psd{y}$ and covariance transfer function $\covtf$.

The Allan variance, the conventional measure of frequency standard instability, can be expressed analogously

\begin{align}
\avar{y}{} &= \frac{1}{2\pi} \int_{0}^{\infty} \psd{y} \atf{} d\omega
\end{align}

\noindent where the transfer function, for ideal Ramsey interrogation, is

\begin{align}
\atf{} &= \frac{2 \sin^4{(\omega\uramsey/2)}}{(\omega \uramsey/2)^2}
\end{align}

\noindent where $\uramsey$ lacks an index because the definition of the Allan variance assumes equal-duration interrogation bins \cite{rutman1978}. The Allan variance calculated via this frequency-domain approach can be compared to its value via the time-domain approach, which consists of finding the variance of the difference between consecutive pairs of measurement outcomes:

\begin{align}\label{eq:allan}
\avar{y}{}&= \frac{1}{2}\langle (\y{k+1} - \y{k})^2 \rangle
\end{align}

\noindent where $\y{k}$ is the $k$th measurement outcome and $\langle\cdots\rangle$ may indicate a time average or an ensemble average, depending on whether $y(t)$ is assumed to be ergodic.

The LLO Allan variance can be found by substituting (\ref{eq:nonrec}) into the definition of the Allan variance (\ref{eq:allan}):
\begin{widetext}
\begin{align}
\avar{k}{LLO} &= \frac{1}{2}\ev{(\yllo{k+1} - \yllo{k})^{2}} \\
&= \frac{1}{2}\evbig{\bigg(\ylo{k+1} - \frac{\gb{k+1}}{\gb{k}} \ylo{k} - \ylo{k} + \frac{\gb{k}}{\gb{k-1}} \ylo{k-1}\bigg)^{2}} \\ 
&= \frac{1}{2}\bigg( \mvar{k+1}{LO} + \bigg(1+\frac{\gb{k+1}}{\gb{k}}\bigg)^{2}\mvar{k}{LO} + \bigg(\frac{\gb{k}}{\gb{k-1}}\bigg)^{2} \mvar{k-1}{LO}\nonumber\\
&+ \frac{2\gb{k}}{\gb{k-1}} \cov{\ylo{k+1}}{\ylo{k-1}} - 2\bigg(1+\frac{\gb{k+1}}{\gb{k}}\bigg) \cov{\ylo{k}}{\ylo{k+1}} - \frac{2(\gb{k} + \gb{k+1})}{\gb{k-1}}\cov{\ylo{k}}{\ylo{k-1}}\bigg)
\end{align}

\end{widetext}

\bibliography{optimizedmeasurement}

\end{document}
%